\documentclass[12pt]{article}

\usepackage{amsfonts}
\usepackage{amssymb}
\usepackage{amscd}      % adapts the commutative diagram macros of AMS-\TeX{}

\def\D{\hbox{D\kern-.73em\raise.25ex\hbox{-}\raise-.25ex\hbox{ }}}
 \def\d{\hbox{d\kern-.33em\raise.75ex\hbox{-}\raise-.75ex\hbox{}}}

\def\GGG{\frak G }
\def\gr3{\GGG\,(\SSS_3)}

\def\gr2{\GGG\,(\SSS_2)}

\def\SSS{\frak S}

\def\al{{\alpha}}
\def\bet{{\beta}}

\def\gam{{\gamma}}

\def\vp{\vspace}
\def\hp{\hspace}
\def\ed{\end{document}}
\def\beq{\begin{equation}}
\def\eeq{\end{equation}}
\def\bea{\begin{eqnarray}}
\def\eea{\end{eqnarray}}
\def\ba{\begin{array}}
\def\ea{\end{array}}
\def\bi{\begin{itemize}}
\def\ei{\end{itemize}}

\def\noi{\noindent}
\def\nn{\nonumber}

 \def\dst{\displaystyle}
\newcommand{\bp}{\noindent\begin{minipage}[c]}
\newcommand{\ep}{\end{minipage}}

\begin{document}
 \baselineskip=11pt

\title{Noncommutative Quantum Mechanics\\ with Path Integral\hspace{.25mm}\thanks{\,
Based on a talk given at the III Summer School in Modern
Mathematical Physics, Zlatibor, 20\,--\,31 August 2004.}}
\author{\bf{Branko Dragovich}\hspace{.25mm}\thanks{\,e-mail address:
dragovich@phy.bg.ac.yu}
\\ \normalsize{Institute of Physics, P.O. Box 57, 11001 Belgrade,}\\  \normalsize{Serbia and Montenegro}
\vspace{2mm} \\ \bf{Zoran Raki\' c}\hspace{.25mm}\thanks{\,e-mail
address: zrakic@matf.bg.ac.yu}
\\ \normalsize{Faculty of Mathematics, P.O. Box 550, 11001 Belgrade,}\\  \normalsize{ Serbia and Montenegro}}

\date{}

\maketitle

%\maketitle

\begin{abstract}
\noi We consider classical and quantum mechanics related to an
additional noncommutativity, symmetric in  position and momentum
coordinates. We show that such mechanical system can be
transformed to the corresponding one which allows employment of
the usual formalism. In particular, we found  explicit connections
between quadratic  Hamiltonians and Lagrangians, in their
commutative and noncommutative regimes. In the quantum case we
give general procedure how to compute Feynman's path integral in
this noncommutative phase space with quadratic Lagrangians
(Hamiltonians). This approach is applied to a  charged particle in
the noncommutative plane exposed to constant homogeneous electric
and magnetic fields.
\end{abstract}

%\clearpage   % delate this line

 {\Large{ }}
\section{ Introduction }

Looking for an approach to solve the problem of ultraviolet
divergences, already in the 1930s Heisenberg  conjectured that
position coordinates might be noncommutative (NC). Snyder
\cite{snyder} was the first who started systematically  to develop
this idea in 1947. An intensive interest to NC quantum theories
emerged after observation of noncommutativity in string theory
with D-branes in 1998. Most of the research has been devoted to
noncommutative (NC) field theory (for reviews, see e.g.
\cite{nekrasov} and \cite{szabo}). NC quantum mechanics (NCQM) has
been also actively investigated. It enables construction of simple
NC models which have relevance to concrete phenomenological
systems and  can be regarded as the corresponding one-particle
nonrelativistic sector of NC quantum field theory. \vspace{2mm}

\noindent We consider here $D$-dimensional NCQM which is based on
the following algebra: \bea\hp{-4mm} [ \hat{x_a},\hat{p_b}] =
i\,\hbar\, ( \delta_{ab} - \frac{1}{4}\, \theta_{ac}\,
\sigma_{cb}), \quad [\hat{x_a},\hat{x_b}] = i\,\hbar\,
\theta_{ab},\quad [\hat{p_a},\hat{p_b}] =i\,\hbar\, \sigma_{ab}  ,
\label{1} \eea where $(\theta_{ab}) = {\Theta}$ and $(\sigma_{ab})
={\Sigma}$ are the antisymmetric matrices with constant elements.
This kind of unusual noncommutativity maintains a symmetry between
canonical variables of the phase space. It also allows  simple
reduction to the usual algebra \bea\hp{-5mm} [
\hat{q_a},\hat{k_b}] = i\,\hbar\, \delta_{ab}, \qquad\quad
[\hat{q_a},\hat{q_b}] = 0,\qquad\quad [\hat{k_a},\hat{k_b}] = 0 ,
\label{2} \eea using the following linear transformations: \bea
\hat{ x_a} = \hat{q_a} - \frac{\theta_{ab}\, \hat{k_b}}{2}\,,
\qquad\qquad \hat{ p_a} = \hat{k_a} + \frac{\sigma_{ab}\,
\hat{q_b}}{2}\, , \label{3} \eea  where summation over repeated
indices is understood.
  In the sequel we
often take $\,\theta_{ab}=\theta\, \varepsilon_{ab}\,$ and
$\,\sigma_{ab}=\sigma\, \varepsilon_{ab},\,$ where \bea \hp{-5mm}
\varepsilon_{ab} = \left\{ \begin{array}{lll}
\hspace{2.5mm}1\,, \quad a<b \\
\hspace{2.5mm}0\,, \quad a=b \\
-1\,, \quad a>b \, .\end{array} \right.  \label{4} \eea  \vp{2mm}

\noindent The  standard Feynman path integral \cite{feynman} \bea
\hp{-4mm} {\mathcal K} (x'',t'';x',t') =\int_{x'}^{x''} \exp \left
( \frac{i}{\hbar}\, \int_{t'}^{t''} L(\dot{q},q, t)\, dt \right
)\, {\mathcal D}q \,, \label{5} \eea where ${\mathcal
K}(x'',t'';x',t')$ is the kernel of the unitary evolution operator
$U (t)$ and $x''=q(t''), \ x'=q(t')$ are end points, we generalize
to the above NCQM with quadratic Lagrangians. Such Lagrangians
contain many important and exactly solvable quantum-mechanical
systems.
 In ordinary quantum mechanics (OQM), Feynman's path integral for quadratic
 Lagrangians  can be
evaluated analytically and obtained result has the form
\cite{steiner} \bea
 \hp{-6mm} {\mathcal K}(x'',t'';x',t') =\frac{1}{(i
h)^{\frac{D}{2}}} \sqrt{\det{\left(-\frac{\partial^2 {\bar
S}}{\partial x''_a
\partial x'_b} \right)}} \exp \left(\frac{2\pi i}{h}\,{\bar
S}(x'',t'';x',t')\right), \label{6} \eea where $ {\bar
S}(x'',t'';x',t')$ is the action for the classical trajectory.
According to (1), (2) and (3), NCQM related to the quantum phase
space $(\hat{p} ,\, \hat{x})$ can be regarded as an OQM on the
standard phase space $(\hat{k} ,\, \hat{q})$ and one can apply
usual path integral formalism.

\noindent A systematic path integral approach to NCQM with
quadratic Lagrangians (Hamiltonians) has been investigated during
the last few years in \cite{dragovich1}, \cite{dragovich2} and
\cite{dragovich3}. In \cite{dragovich1} and \cite{dragovich2},
general connections between quadratic Lagrangians and Hamiltonians
for standard and $\theta \neq 0$, $\sigma =0$ noncommutativity are
established, and this formalism was applied to the two-dimensional
NC particle in a constant field and to the NC harmonic oscillator.
Paper \cite{dragovich3} presents generalization of papers
\cite{dragovich1} and \cite{dragovich2} on noncommutativity (1).
The present article reviews general formalism of \cite{dragovich3}
and contains its application to a charged particle in NC plane
exposed to homogeneous electric and magnetic fields.

\section{Quadratic Lagrangians and Hamiltonians
Related to Noncommutative Phase Space}

Let the most general  quadratic Lagrangian for a $D$-dimensional
system with position coordinates $x^T = (x_1 , x_2 , \cdots ,
x_D)$ be \bea \hp{-7mm} L(\dot{x}, x,t) = \frac{1}{2}\,\left(
\dot{x}^T  \alpha\, \dot{x} + \dot{x}^T  \beta\, x  + x^T \beta^T
\dot{x} + x^T  \gamma\, x \right)+ \delta^T \dot{x} + \eta^T x +
\phi \,, \label{7} \eea where coefficients of the $D\times D$
matrices $\alpha =((1+\delta_{ab})\, \alpha_{ab}(t)), $ \linebreak
$ \beta =(\beta_{ab}(t)),\ \gamma =((1+\delta_{ab} )\,
\gamma_{ab}(t)),$ $D$-dimensional vectors $ \delta
=(\delta_{a}(t)),$ \ $\eta =(\eta_{a}(t))$ and a scalar $\phi
=\phi(t)$ are some analytic functions of  the time $t$. Matrices
$\alpha$ and $\gamma$ are symmetric,  $\alpha$ is nonsingular
$(\det\alpha \neq 0)$ and index ${}^T$ denotes transposition.
\vspace{2mm}

\noindent The  Lagrangian (7) can be rewritten in the  more
compact form: \bea L(X,t) = \frac{1}{2}\, X^T \, M\, X + N^T \, X
+ \phi , \label{8} \eea
 where $2D\times 2D$ matrix $M$ and $2D$-dimensional vectors
$X,\, N$ are defined as \bea \hp{-7mm}
 {M } = \left(
\begin{array}{ccc}
 \alpha & \beta  \\
 \beta^T & \gamma \end{array}
\right)\, , \qquad  X^T = (\dot{x}^T\, ,\,  x^T) \, , \qquad N^T =
(\delta^T\, ,\,  \eta^T) . \label{9} \eea Using the equations $
p_a = {\partial L \over \partial \dot x_a},$ one finds $ {\dot x }
= {\alpha^{-1}}\, ({p} - {\beta}\, {x } - \delta ). $ Since the
function $\dot{x}$ is linear in $p$ and $x$, the corresponding
classical Hamiltonian  $ H(p,x,t)= {p}^T\, \dot {x } -
L(\dot{x},x,t)$ becomes also quadratic, i.e. \bea\hp{-9mm}
H(p,x,t) = \frac{1}{2}\,\left( {p}^T \!\hp{-.2mm} A\hp{.2mm} {p} +
p^T\hp{-.5mm} B\hp{.2mm} x + x^T\hp{-.2mm}
 B^T\hp{-.2mm} p  +  x^T\hp{-.2mm}  C\hp{.2mm} x \right) + D^T\hp{-.2mm} p + E^T\hp{-.2mm} x + F
, \label{10} \eea where: \bea \hp{-8mm} \begin{array}{ll}  {A } =
{\alpha}^{-1}, \hspace{1.7cm} {B } =-\, {\alpha}^{-1}\, {\beta},
\hspace{1.7cm} {C} =
{\beta}^T\, {\alpha}^{-1}\, {\beta} - {\gamma } , \vp{1mm} \\
  {D} =- \, {\alpha }^{-1}\,  \delta, \hspace{1.0cm}  {E } =
\beta^T \, \alpha^{-1}\,  \delta - \eta , \hspace{1.0 cm} {F} =
\displaystyle{{1\over 2}}\, {\delta}^T \, {\alpha}^{-1} {\delta}\,
- {\phi } \, .
\end{array}  \label{11} \eea

\noindent Due to the symmetry of matrices $\alpha$ and $\gamma$
one can easily see that matrices $\,A = ((1+\delta_{ab})\,
A_{ab}(t))\,$ and $\,C = ((1+\delta_{ab})\, C_{ab}(t))\,$ are also
symmetric ($A^T = A ,\, \, C^T = C$). The nonsingular $(\det
{\alpha}\neq 0)$ Lagrangian $L(\dot{x},x,t)\, $   implies
nonsingular $ (\det{ A}\neq 0) $ Hamiltonian $H(p,x,t) $. Note
that the inverse map, i.e. $H \to L$, is given by the same
relations (11). \vspace{2mm}

\noi The Hamiltonian (10) can be also presented in the compact
form \bea H(\Pi,t) = \frac{1}{2}\; \Pi^T  {\mathcal M}\; \Pi +
{\mathcal N}^T \,  \Pi   + F , \label{12} \eea
 where matrix ${\mathcal M}$ and
vectors $\Pi,\, {\mathcal N}$ are \bea \hp{-7mm}   {\mathcal M } =
\left(
\begin{array}{ccc}
 A & B  \\
 B^T & C \end{array}
\right)\, , \qquad  \Pi^T = (p^T\, ,\,  x^T) \, , \qquad {\mathcal
N}^T = (D^T\, ,\,  E^T)\, . \label{13} \eea

\noi One can easily show that \bea  {\mathcal M}= \sum_{i =1}^3
\Upsilon_i^T(M)\,M\, \Upsilon_i(M) , \label{14} \eea where \bea
\hp{-7mm} \ba{l} \Upsilon_1(M)=\left( \ba{lr} \al^{-1} &0\\0 &
-I\ea\right), \qquad \qquad
\Upsilon_2(M)=\left( \ba{lr} 0\  &  \alpha^{-1} \beta\\
0 & 0\ea\right), \vp{4mm} \\
 \Upsilon_3(M)=\left( \ba{lr} 0\  &0\\0 & \
 i\sqrt{2}\,I\ea\right) ,\ea \label{15}
 \eea
 and $I$ is $D \times D$ unit matrix.
 One has also  ${\mathcal N}= Y(M) \, N,  $
 where
 \bea \hp{-7mm} Y (M)= \left(\ba{rr}
-\,\al^{-1} & 0  \\ \bet^T\,\al^{-1}& -I  \ea\right) = -\Upsilon_1
(M) + \Upsilon^T_2 (M) + i\, \sqrt{2}\,\, \Upsilon_3 (M)\,,
\label{16} \eea and $F = N^T\, Z(M)\, N - \phi ,$ where \bea
\hp{-7mm} Z(M ) = \left(\ba{ll} \frac{1}{2}\, \al^{-1} & 0  \\ 0 &
0 \ea\right) =\frac{1}{2}\, \Upsilon_1 (M) - \frac{i}{2\sqrt{2}}\,
\Upsilon_3 (M)\, . \label{17}\eea We have shown that Hamiltonian
compact quantities ${\mathcal M}, {\mathcal N}$ and $F$ can be
related to the corresponding Lagrangian ones $M, N$ and $\phi$
using auxiliary matrices $\Upsilon_1 (M), \Upsilon_2 (M)$ and
$\Upsilon_3 (M)$. \vp{2mm}

 \noi Eqs. (3) can be rewritten in the compact form as
\bea\hp{-8mm} \hat{\Pi} = \Xi \,\, \hat{K} ,\quad \qquad \Xi =
\left(
\begin{array}{ccc}
 I & \frac12\,\, {\Sigma}  \\
- \frac12\,\, {\Theta} & I \end{array} \right) , \quad\qquad
\hat{K}= \left(
\begin{array}{ccc}
 \hat{k}  \\
 \hat{q} \end{array}
\right) .\  \label{18} \eea

\noi Since Hamiltonians depend on canonical variables, the
transformation  (18) leads to the transformation of Hamiltonians
(10) and (12). To this end, let us quantize the Hamiltonian (10)
and it easily becomes \bea \hp{-8mm} H(\hat{p},\hat{x},t) =
\frac{1}{2}\, ( {\hat{p}}^T\! A\, \hat{p} + \hat{p}^T \! B\,
\hat{x} + \hat{x}^T \! B^T\! \hat{p} + \hat{x}^T \! C\, \hat{x} )
+ D^T\! \hat{p} + E^T\! \hat{x} + F \label{19} \eea because (10)
is already written in the Weyl symmetric form. \vp{2mm}

\noi Performing linear transformations (3) in the above
Hamiltonian we again obtain quadratic quantum Hamiltonian \bea\nn
\hp{-9mm} H_{\theta\sigma}(\hat{k},\hat{q},t) &=&
\frac{1}{2}\,\left( {\hat{k}}^T \hp{-.5mm} A_{\theta\sigma}\,
\hat{k} + \hat{k}^T \hp{-.5mm} B_{\theta\sigma}\, \hat{q} +
\hat{q}^T \hp{-.5mm} B^T_{\theta\sigma}\, \hat{k} + \hat{q}^T
\hp{-.5mm} C_{\theta\sigma}\, \hat{q} \right) \\ && \label{20}
 +\ D^T_{\theta\sigma}\, \hat{k} + E^T_{\theta\sigma}\, \hat{q} +
F_{\theta\sigma}\, ,   \eea where    \bea \hp{-8mm} \ba{ll}
\displaystyle{{A}_{\theta\sigma} = {A} - {1\over 2}\,\, {B}\, {
\Theta} + {1\over 2}\,\, {\Theta}\, {B}^T -{1\over 4}\,\,
{\Theta}\, {C}\, { \Theta} ,} & \hp{5mm}
\displaystyle{{D}_{\theta\sigma} = {D} + {1\over
2}\,\, {\Theta}\, {E}}, \vp{2mm} \\
\displaystyle{{B}_{\theta\sigma} = {B} + \frac{1}{2}\, {\Theta}\,
{C} + \frac{1}{2}\, A\, \Sigma + \frac14\,\, \Theta\, B^T \,
\Sigma} , & \hp{5mm} \displaystyle{{E}_{\theta\sigma} = {E}-
\frac 12\, \Sigma\, D} ,  \vp{2mm}\\
\displaystyle{{C}_{\theta\sigma} = {C} - {1\over 2}\,\Sigma\, {B}
+ {1\over 2}\, B^T \, {\Sigma} -{1\over 4}\, {\Sigma}\, {A}\,
\Sigma}, & \hp{5mm} \displaystyle{{F}_{\theta\sigma} = {F}} .\ea
 \label{21}
\eea Note that for the nonsingular Hamiltonian $
H(\hat{p},\hat{x},t)$ and for sufficiently small $\theta_{ab}$ the
Hamiltonian $ H_{\theta\sigma}(\hat{k},\hat{q},t)$ is also
nonsingular. It is worth noting that $A_{\theta\sigma}$ and
$D_{\theta\sigma}$ do not depend on $\sigma$, as well as
$C_{\theta\sigma}$ and $E_{\theta\sigma}$ do not depend on
$\theta$. Classical analogue of (20) maintains the same form
\bea\nn \hp{-7mm} H_{\theta\sigma}(k,q,t) &=& \frac{1}{2}\, (
k^T\hp{-.5mm} A_{\theta\sigma}\, {k} + {k}^T \hp{-.5mm}
B_{\theta\sigma}\, {q} + {q}^T\hp{-.2mm} B^T_{\theta\sigma}\, {k}
+ {q}^T\hp{-.2mm} C_{\theta\sigma}\, {q} ) \\ && \nn  +\
D^T_{\theta\sigma}\, {k} + E^T_{\theta\sigma}\, {q} +
F_{\theta\sigma}\,.\eea

\noi In the more compact form, Hamiltonian (20) is \bea\hp{-7mm}
 \hat
H_{\theta\sigma}(\hat{K},t) = \frac{1}{2}\,\, \hat K^T \,
{\mathcal M}_{\theta\sigma}\; \hat K + {\mathcal
N}_{\theta\sigma}^T \; \hat K + F_{\theta\sigma} , \label{22} \eea
where $2D\times 2D$ matrix ${\mathcal M}_{\theta\sigma}$ and
$2D$-dimensional vectors $\hat{K},\, {\mathcal N}_{\theta\sigma}$
are \bea\hp{-9mm}  {\mathcal M}_{\theta\sigma} = \left(
\begin{array}{ccc}
 A_{\theta\sigma} & B_{\theta\sigma}  \\
 B_{\theta\sigma}^T & C_{\theta\sigma} \end{array}
\right) , \ \ \quad  \hat{K}^T = (\hat{k}^T\,, \,  \hat{q}^T)  ,
\quad\ \ {\mathcal N}_{\theta\sigma}^T = (D_{\theta\sigma}^T\,, \,
E_{\theta\sigma}^T) . \label{23} \eea From (12), (18) and (22) one
can find connections between ${\mathcal M}_{\theta\sigma},\,
{\mathcal N}_{\theta\sigma} , \, { F}_{\theta\sigma} $ and
${\mathcal M},\, {\mathcal N}, \, F$, which are given by the
following relations: \bea\hp{-10mm}   {\mathcal
M}_{\theta\sigma}=\Xi^T\,{\mathcal M}\,\, \Xi \,, \quad\qquad
{\mathcal N}_{\theta\sigma}=\Xi^T\,{\mathcal N} , \qquad\quad
{F}_{\theta\sigma} = F.  \label{24}  \eea The corresponding
Schr\"odinger equation is \bea\hp{-4mm} i\, \hbar \frac{\partial
\Psi (q, t)}{\partial t} = {{H}_{\theta\sigma} (\hat{k}, q, t)}\,
\Psi (q, t) \label{25} \eea and this approach has been mainly
exploited to analyze dynamical evolution in NCQM . \vp{2mm}

\noi To compute a path integral, which is a basic instrument in
Feynman's approach to quantum mechanics, one can start from its
Hamiltonian formulation on the phase space. However, such path
integral on a phase space can be reduced to the Lagrangian path
integral on configuration space whenever Hamiltonian is a
quadratic polynomial with respect to momentum $k$ (see, e.g.
\cite{dragovich2}). Hence, we would like to have the corresponding
classical Lagrangians  related to the Hamiltonians (20) and (22).
Using equations $ \dot{q}_a = \frac{\partial
H_{\theta\sigma}}{\partial p_a} $ which give $ k =
A_{\theta\sigma}^{-1}\, (\dot{q} - B_{\theta\sigma}\, q -
D_{\theta\sigma}) $ we can pass from Hamiltonian (20) to the
corresponding Lagrangian by relation $ L_{\theta\sigma}
(\dot{q},q,t) = k^T \dot{q}  - H_{\theta\sigma} (k,q,t) . $ Note
that coordinates $q_a$ and $x_a$ coincide when $\theta =\sigma
=0$. Performing necessary computations we obtain \bea \nn
\hp{-8mm} L_{\theta\sigma}(\dot{q}, q,t) &=& \frac{1}{2}\,\left(
\dot{q}^T  \alpha_{\theta\sigma}\, \dot{q} + \dot{q}^T
\beta_{\theta\sigma}\, q  +  q^T  \beta^T_{\theta\sigma}\,
\dot{q}  +  q^T  \gamma_{\theta\sigma}\, q \right)  \\
 && +\  \delta^T_{\theta\sigma}\, \dot{q} + \eta^T_{\theta\sigma}\, q
+ \phi_{\theta\sigma} , \label{26} \eea or in the  compact form:
\bea\hp{-8mm} L_{\theta\sigma}(Q,t) = \frac{1}{2}\; Q^T \,
M_{\theta\sigma}\; Q+ N_{\theta\sigma}^T \; Q +
\phi_{\theta\sigma} , \label{27} \eea  where \bea \hp{-7mm}
{M_{\theta\sigma}} = \left(
\begin{array}{ccc}
 \alpha_{\theta\sigma} & \beta_{\theta\sigma}  \\
 \beta_{\theta\sigma}^T & \gamma_{\theta\sigma} \end{array}
\right)\, , \quad   Q^T = (\dot{q}^T\,, \,  q^T) \, , \quad N^T =
(\delta_{\theta\sigma}^T\,, \,  \eta_{\theta\sigma}^T). \label{28}
\eea Then the connection between ${M_{\theta\sigma}},\,
{N_{\theta\sigma}},\, \phi_{\theta\sigma} \, $ and $\, M,\,
N,\,\phi$ are given by the following relations: \bea \hp{-9mm}
\ba{l} M_{\theta\sigma} = \sum\limits_{i,j=1}^3 \, \Xi_{ij}^T\, \,
M\, \,\Xi_{ij},\qquad \qquad\hp{1.6mm}
  \Xi_{ij}=\Upsilon_i(M)\,\,\Xi\,\,\Upsilon_j(\mathcal{M}_{\theta\sigma})
, \vp{3mm} \\
 N_{\theta \sigma} = Y (\mathcal{M}_{\theta\sigma})\, \Xi^T\,
Y(M)\, N, \quad \quad \phi_{\theta \sigma} =
\mathcal{N}_{\theta\sigma}^T \, Z (\mathcal{M}_{\theta\sigma}) \,
\mathcal{N}_{\theta\sigma} - F  \, .\ea  \label{29} \eea

 \noi In more detail, the connection between  coefficients of the
 Lagrangians $L_{\theta\sigma}$
 and  $L$ is given by the relations:
 \bea \hp{-7mm}\ba{l} {\al}_{\theta\sigma}  = \big[\,
{\al}^{-\,1} - {\frac 12} \, (\Theta\, {\bet}^{T}\, {\al}^{-\,1}-
{\al}^{-\,1}\, {\bet}\, \Theta) -\frac14\,\Theta\,( {\bet}^{T}\,
{\al }^{-\,1}\, {\bet} - \gam)
\,\Theta \, \big]^{-\,1}\,, \vp{2mm}  \\
  {\bet}_{\theta\sigma}  =  {\al}_{\theta\sigma}\,
 \big( {\al}^{-\,1}\,
 {\bet}  - {\frac 12}\,( {\al}^{-\,1}\,\Sigma -\Theta\, {\gam}+
  \Theta\, {\bet}^{T}\, {\al}^{-\,1}\,{\bet})
  +\frac14\,\Theta\,{\bet}^{T}\,{\al}^{-\,1}\,\Sigma\big)\,, \vp{2mm} \\
 {\gam}_{\theta\sigma} = \gamma +  \bet_{\theta\sigma}^T \,
\al_{\theta\sigma}^{-\,1} \, \bet_{\theta\sigma} -  {\bet }^{T}\,
{\al}^{-\,1}\, {\bet}  + {\frac 14}\, \Sigma\,{\al}^{-\,1}\,
\Sigma\,\vp{1mm} \\ \hp{11mm}  -\ {\frac
12}\,(\Sigma\,{\al}^{-\,1}\,\bet -{\bet }^{T}\,
{\al}^{-\,1}\,\Sigma ) \,, \vp{2mm} \\
 {\delta}_{\theta\sigma}  =
{\al}_{\theta\sigma}\,\big({\al}^{-\,1}\,{\delta}+\frac
12\,\,(\Theta\,\eta- \Theta\,{\bet}^{T}\,
{\al}^{-\,1}\,{\delta}) \big)\,, \vp{2mm}  \\
 {\eta }_{\theta\sigma} = \eta + \bet_{\theta\sigma}^T \,
\al_{\theta\sigma}^{-\,1} \, \delta_{\theta\sigma} -
{\bet}^{T}\,{\al}^{-\,1}\, {\delta}  - \frac 12\,\,
\Sigma\,{\al}^{-\,1}\,{\delta}\,,\vp{2mm}  \\
 {\phi }_{\theta\sigma}  =  \phi + \frac 12\,\,
{\delta}_{\theta\sigma}^T\,  \al_{\theta\sigma}^{-\,1} \,
\delta_{\theta\sigma} - \frac 12\,\, \delta^T\,
{\al}^{-\,1}\,{\delta} \,. \ea
 \label{30} \eea
Note that $\alpha_{\theta \sigma}, \,  \delta_{\theta \sigma} $
and $ \phi_{\theta \sigma}$ do not depend on $\sigma$. \vp{3mm}

\section{Noncommutative Path Integral}

If we know Lagrangian (7) and algebra (1) we can obtain the
corresponding effective Lagrangian (26) suitable for the path
integral in NCQM. Exploiting the Euler-Lagrange equations \bea\nn
\frac{\partial L_{\theta\sigma}}{\partial q_a} -\frac{d}{dt}
\frac{\partial L_{\theta\sigma}} {\partial{\dot q}_a} =0 \eea one
can obtain the classical trajectory $q_a =q_a (t)$ connecting  end
points $x' = q(t')$ and $x''= q(t'')$, and the corresponding
action \bea\nn {\bar S}_{\theta\sigma} (x'',t'';x',t')
=\int_{t'}^{t''} L_{\theta\sigma} (\dot{q}, q,t)\, dt.\eea Path
integral in NCQM is a direct analogue of (5) and its exact
expression in the form of quadratic actions ${\bar
S_{\theta\sigma}}(x'',t'';x',t')$ is \bea && {\mathcal
K}_{\theta\sigma}(x'',t'';x',t') = \frac{1}{(i h)^{\frac{D}{2}}}
\sqrt{\det{\left(-\frac{\partial^2 {\bar
S_{\theta\sigma}}}{\partial x''_a\,
\partial x'_b} \right)}}  \exp \left(\frac{2\,\pi\, i}{h}\,{\bar
S_{\theta\sigma}}(x'',t'';x',t')\right).\nn \\ &&  \label{31} \eea
\vp{3mm}

 \section{A charged particle in a noncommutative plane exposed to
 homogeneous electric and magnetic fields }

As an example to illustrate many features of the above formalism
we consider a particle of a charge $e > 0$ moving in a plane
$(x_1, x_2 )$ with noncommutativity  parameters $\theta $ and
$\sigma$. Let this particle be also under the influence of  a
constant electric field ${\mathcal E}$ along coordinate $x_1$ and
a constant magnetic field ${\mathcal B}$ perpendicular to the
plane and oriented along the axis $-x_3$. It is suitable to start
from the nonrelativistic Hamiltonian \bea H(p, x) = \frac{1}{2\,
m} \, \Big[ (p_1 - e\, {\mathcal A}_1)^2 + (p_2 - e\, {\mathcal
A}_2)^2 \Big] + e \, \varphi ,  \label{32} \eea
 where    ${\mathcal A}_1 =\frac{{\mathcal B} }{2}\, x_2\, , \, \, {\mathcal
A}_2 =-\,\frac{{\mathcal B} }{2}\, x_1$ and $\varphi = - {\mathcal
E}\, x_1$ .  Using the inverse  map of (11) we get the
corresponding Lagrangian \bea\hp{-5mm}
 L (\dot{x}, x) = \frac{m}{2}\, (\dot{x}_1^2 +
\dot{x}_2^2 )  + \frac{e\, {\mathcal B}}{2}\, (\dot{x}_1 x_2 -
\dot{x}_2 x_1) + e \hp{.7mm} {\mathcal E}\hp{.3mm} x_1 .
\label{33} \eea Employing formulas  (21) and (30)  one obtains
Hamiltonian \bea H_{\theta\sigma} (k, q) &=& \frac{1}{2\hp{.3mm}
\mu}\, (k_1^2 + k_2^2) - \frac{\lambda}{2\hp{.3mm} \mu}\,
(k_1\hp{.3mm} q_2 -
k_2\hp{.5mm} q_1) + \frac{\lambda^2}{8\hp{.3mm} \mu}\, (q_1^2  + q_2^2) \nn \\
&&  +\ \frac{\theta\hp{.3mm} e\hp{.6mm} {\mathcal E}}{2}\, k_2 -
e\hp{.6mm} {\mathcal E}\hp{.2mm} q_1 \label{34} \eea and
Lagrangian
$$ L_{\theta\sigma} (\dot{q}, q) = \frac{\mu}{2} \, (\dot{q}_1^2 +
\dot{q}_2^2) + \frac{\lambda}{2} \,  (\dot{q}_1\hp{.3mm} q_2 -
\dot{q}_2\, q_1)  - \frac{\mu\hp{.3mm} \theta\hp{.2mm} e\hp{.6mm}
{\mathcal E}}{2}  \, \dot{q}_2 + \nu_0\, q_1 + \frac{\mu\hp{.5mm}
\theta^2\hp{.2mm} e^2\hp{.5mm} {\mathcal E}^2}{8} \, , \eqno{(35)}
$$
 where \setcounter{equation}{35}
\bea \hp{-4mm}
 \mu = \frac{m}{\big( 1- \frac{\theta\, e\,
{\mathcal B}}{4} \big)^2}\, \, , \quad \quad \lambda = \frac{e\,
{\mathcal B} - \sigma}{ 1- \frac{\theta\, e\, {\mathcal
B}}{4}}\,\,, \qquad \nu_0=e\, {\mathcal{E}}\,\Big(1+\frac{\theta\,
\lambda}{4}\Big)\,.   \label{36}\eea
 The
above Hamiltonian and Lagrangian are related to the dynamics in
noncommutative phase space, where noncommutativity is
characterized by parameters $\theta$ and $\sigma$. \vp{2mm}

\noi  The Lagrangian given by (35) implies the Euler-Lagrange
equations, \bea \hp{-5mm} \displaystyle{\mu\,\ddot{q}_1+
\lambda\,\dot{q}_2 =\nu_0,} \qquad \qquad
\displaystyle{\mu\,\ddot{q}_2- \lambda\,\dot{q}_1 =0\,.}
\label{37} \eea
 One can transform the system (37)  to
\bea \ba{l} \displaystyle{\mu^2\,{q}_1^{(3)}+ \lambda^2\,
{q}_1^{(1)}=0,} \hspace{15mm} \displaystyle{\mu^2\, {q}_2^{\,(3)}+
\lambda^2\,{q}_2^{(1)}-\lambda\, \nu_0=0\,.} \ea \label{39} \eea
 The
solution of the equations (38) has the following form \bea
\hspace{-16mm}  \ba{ll}  {q}_1(t) = C_1 + C_2 \cos
(\eta\, t) + C_3 \sin (\eta\, t)\,,   \vp{2mm}\\
{q}_2(t) = D_1 + D_2\, \cos (\eta\, t) + D_3 \sin (\eta\, t)+
\dst{\frac{\nu_0}{\lambda}}\,t\,, \qquad \eta=
\dst{\frac{\lambda}{\mu}}\,. \ea \label{39} \eea Imposing coupled
differential equations (37) on $q_1$ and $q_2$ , we obtain the
following connection between constants: $ C_2=D_3,\quad
D_2=-C_3\,. $ The unknown constants $C_1,D_1,C_3$ and $D_3$  can
be fixed from initial conditions given by \bea \hp{-10mm}
q_1(0)=x_1',\qquad q_1(T)=x_1'', \qquad q_2(0)=x_2',\qquad
q_2(T)=x_2''.  \label{40} \eea The corresponding constants are :
\bea
 \hp{-10mm} \ba{l} \displaystyle{C_1=
\frac{x_1'+x_1''}{2}+
\frac{x_2'-x_2''}{2}\,\cot\!\Big[\hp{.2mm}\frac{T\hp{.3mm}\eta}{2}\hp{.2mm}\Big]+
\frac{T\hp{.3mm}\nu_0}{2\hp{.3mm}\lambda}\,
\cot\!\Big[\hp{.2mm}\frac{T\,\eta}{2}\hp{.2mm}\Big], } \vp{2mm}\\
  \displaystyle{C_3=
\frac{-x_2'+x_2''}{2}-
\frac{x_1'-x_1''}{2}\,\cot\!\Big[\hp{.2mm}\frac{T\hp{.3mm}\eta}{2}\hp{.2mm}\Big]-
\frac{T\hp{.3mm}\nu_0}{2\hp{.3mm}\lambda}\,, }
  \vspace{2mm} \\
\displaystyle{D_1= \frac{x_2'+x_2''}{2}-
\frac{x_1'-x_1''}{2}\,\cot\!\Big[\hp{.2mm}\frac{T\hp{.3mm}\eta}{2}\hp{.2mm}\Big]-
\frac{T\hp{.3mm}\nu_0}{2\,\lambda}\,, }
  \vspace{2mm} \\
  \displaystyle{D_3=
\frac{x_1'-x_1''}{2}-
\frac{x_2'-x_2''}{2}\,\cot\!\Big[\hp{.2mm}\frac{T\hp{.3mm}\eta}{2}\hp{.2mm}\Big]-
\frac{T\hp{.3mm}\nu_0}{2\hp{.3mm}\lambda}\,
\cot\!\Big[\hp{.2mm}\frac{T\hp{.3mm}\eta}{2}\hp{.2mm}\Big] \,.}
\ea \label{41} \eea

\noi Inserting the above expressions for constants (41) into
 (39) we obtain solutions of Euler--Lagrange equations (37).
 For these solutions and their time derivatives  we
find the following expression for Lagrangian (35): \bea\hp{-7mm}
\ba{l} L_{\theta\sigma}(\dot q,q) =
\dst{\frac{1}{8\hp{.3mm}\lambda^2\hp{.3mm}\mu}}\Big(\mu\big(
4\hp{.3mm}C_1\hp{.3mm}\lambda^2 \hp{.3mm} \nu_0
+\mu\hp{.3mm}(-2\hp{.3mm}\nu_0\ +\theta\hp{.3mm}
\lambda\hp{.3mm} e\hp{.7mm} {\mathcal{E}})^2\big)+4\hp{.3mm} \lambda^2 \vp{1.0mm} \\
\hp{3mm} \times \, \Big( \big(
C_3\hp{.3mm}\lambda\hp{.3mm}(D_1\hp{.3mm}\lambda+\nu_0\hp{.3mm}t)
 - D_3\hp{.3mm}(C_1\hp{.3mm}\lambda^2 -3\hp{.4mm}\mu\hp{.3mm}\nu_0  +
\theta\hp{.3mm}\lambda\hp{.4mm}\mu\hp{.3mm}
e\hp{.7mm}{\mathcal{E}})\big)\hp{.3mm}\cos[\hp{.2mm}\eta\hp{.3mm}t\hp{.2mm}]
\vp{1.0mm}\\
 \hp{3mm} -\, \big( (C_1\hp{.3mm}C_3 + D_1\hp{.3mm}D_3)\hp{.3mm} \lambda^2
 + D_3\hp{.3mm}\lambda\hp{.3mm}\nu_0\hp{.3mm}t -3\hp{.4mm}C_3\hp{.3mm}\mu\hp{.3mm}\nu_0 + C_3\hp{.3mm}
\theta\hp{.3mm}\lambda\hp{.4mm}\mu\hp{.3mm}
e\hp{.7mm}{\mathcal{E}}\big)
 \hp{.3mm}\sin[\hp{.2mm}\eta\hp{.3mm}t\hp{.2mm}]\Big)\!\Big)\,. \ea \label{42}\eea
 Using (42), we finally compute the
corresponding action \bea \label{43}  \hp{-7mm} \ba{l} {\bar
S}_{\theta\sigma} (x'',T;x',0) = \hp{-.4mm}\dst{\int\limits_{0}^T
\! L_{\theta\sigma}(\dot{q},q)\, d\hspace{.1mm}t} =
 \dst{\frac{\lambda}{2}\hp{-.2mm}\Big(\hp{-.2mm}x'_2\,x''_1-x'_1\,x''_2\hp{-.2mm} \Big)+
 \frac{\nu_0\hp{.2mm} T}{2}\hp{-.2mm}\Big(\hp{-.2mm}x'_1+x''_1\hp{-.2mm}\Big)} \vp{1.5mm} \\
\hp{7mm}  \dst{+\
\frac{\mu\hp{.2mm}\nu_0}{\lambda}\hp{-.2mm}\Big(\hp{-.2mm}x''_2-x'_2\hp{-.2mm}\Big)+\frac{\theta\hp{.2mm}\mu\hp{.2mm}
e\hp{.6mm}
{\mathcal{E}}}{2}\hp{-.2mm}\Big(\hp{-.2mm}x'_2-x''_2\hp{-.2mm}\Big)+
  \frac{\mu\hp{.2mm}T}{8}\hp{-.2mm}\Big(\hp{-.2mm}-4\,\frac{\nu_0^2}{\lambda^2}+
 \theta^2\hp{.2mm}e^2\hp{.3mm}{\mathcal{E}}^2\hp{-.2mm}\Big)} \vp{2.0mm}  \\
 \hp{7mm} \dst{+\ \frac{(x'_1-x''_1)^2\,\lambda^2 + ((x'_2-x''_2)\,\lambda+T\,
\nu_0)^2}{4\,\lambda} \,
\cot\!\Big[\hp{.2mm}\frac{\eta\,t}{2}\hp{.2mm}\Big] \, .} \ea
\eea Accordingly
 we obtain
\bea \hp{-8mm} \det{\left(\!-\frac{\partial^2 {\bar
S_{\theta\sigma}}}{\partial x''_a\,
\partial x'_b}\right)} =\frac{\lambda^2}{4\,
\sin^2\big[\frac{\lambda\,T}{2}\big]}\,\,, \label{44}\eea  and
finally \bea\hp{-5mm}
  {\cal K}_{\theta\sigma} (x'',T;x',0) =
 \frac{|\hp{.4mm}\lambda\,\big|}
 {2\,i\, h\,\big|\!\sin\big[\frac{\lambda\,T}{2}\big]\hp{-.1mm}\big|}\,
 \exp\left(
 \frac{2\pi i}{h}\, \bar{S}_{\theta\sigma} (x'',T;x',0)   \right)
,  \label{45} \eea where $\bar{S}_{\theta\sigma} (x'',T;x',0)$ is
given by (43).

\section{Discussion and Concluding Remarks}

 Let us mention  that taking $\,\sigma=0 ,\, \,  \theta=0\,$ in
the above formulas we recover expressions for the Lagrangian
$\,L(\dot{q}, q),\,$ action $\,\bar{S} (x'',T;x',0)\, $ and
probability amplitude $\,{\mathcal K} (x'',T;x',0)\,$  of the
ordinary commutative case. \vp{2mm}

\noi Note that a similar path integral approach with $\,\sigma
=0\,$
 has been considered
in the context of the Aharonov-Bohm effect, the Casimir effect, a
quantum system in a rotating frame, and the Hall effect
(references on these and some other related subjects can be found
in \cite{dragovich1,dragovich2,dragovich3}). Our investigation
contains all quantum-mechanical systems with quadratic
Hamiltonians (Lagrangians) on NC phase space given by (1).\vp{2mm}

\noi On the basis of the expressions presented in this article,
there are many possibilities to discuss noncommutative
quantum-mechanical systems with respect to various values of
noncommutativity parameters $\,\theta\,$ and $\,\sigma.\,$ We will
discuss only some aspects of the above two-dimensional model.
\vp{2mm}

\noi Since \bea \hp{-9mm} [\hp{.2mm}\hat{\pi}_a ,
\hat{\pi}_b\hp{.2mm}] = -\hp{.2mm} i\hp{.2mm} \hbar\, ( e\hp{.3mm}
{\mathcal B} - \sigma) \Big(1 - \frac{e\hp{.3mm} {\mathcal B}
\hp{.3mm}\theta}{4} \Big)\, \varepsilon_{ab} ,      \label{46}
\eea where $ \, \hat{\pi}_a = \hat{p}_a -e \hp{.2mm}\hat{\mathcal
A}_a, $ a charged particle in a plane with the perpendicular
background magnetic field $\,{\mathcal B}\,$ and phase space
noncommutativity (1) lives under an extended momentum
noncommutativity depending on
 $\,  -\hp{.2mm} ( e\hp{.3mm} {\mathcal B} - \sigma)
 \big(1 - \frac{e\hp{.3mm} {\mathcal B}\hp{.2mm} \theta}{4}
\big)$. Note that $\,p\,$ is a canonical momentum and $\,\pi\,$ is
the corresponding physical one.\vp{2mm}

\noi Using annihilation and creation operators \bea\hp{-10mm} a =
\frac{\hat{\pi_1} - i\hp{.2mm}
\hat{\pi_2}}{\sqrt{\hp{.2mm}2\hp{.3mm} \hbar\hp{.3mm} m\hp{.3mm}
{\omega}}}\, , \qquad\qquad a^\dagger  = \frac{\hat{\pi_1} +
i\hp{.2mm} \hat{\pi_2}}{\sqrt{\hp{.2mm}2\hp{.3mm} \hbar\hp{.3mm} m
\hp{.3mm}{\omega}}}\,, \label{47} \eea where the frequency \bea
\hp{-9mm}{\omega} = \frac{( e\hp{.3mm} {\mathcal B} - \sigma)
\big( 1 - \frac{e\hp{.3mm} {\mathcal B}\hp{.2mm} \theta}{4}
\big)}{m}\,, \label{48} \eea one can write the Hamiltonian (32)
with NC phase space (1) and the electric field $\,{\mathcal E} =
0\,,$ in the harmonic oscillator form \bea \hp{-10mm}\hat{H} =
{\omega}\hp{.3mm} \hbar \hp{.3mm} \Big(\hp{.3mm}
a^\dagger\hp{.3mm} a  + \frac{1}{2} \hp{.3mm} \Big) \label{49}
\eea with energy levels $\,E_n = \omega\hp{.3mm} \hbar \hp{.3mm}
(n + \frac{1}{2}) $. This is an extended Landau problem which
reduces to the standard one if $\,\sigma = \theta = 0.\,$\vp{2mm}

\noi The frequency $\,\omega\,$ can tuned by magnetic field
$\,{\mathcal B}.\,$ In particular, $\,\omega = 0\,$ if
$\,{\mathcal B} =\frac{\sigma}{e}\,$ or $\,{\mathcal B} =
\frac{4}{e\hp{.3mm} \theta}.\,$ Note that $\,\omega_{{\mathcal B}
\to 0} = - \frac{\sigma}{m}\,$ and $\,\omega_{{\mathcal B} \to
\infty} = \frac{e^2\hp{.2mm} {\mathcal B}^2\hp{.2mm}
\theta}{4\hp{.3mm} m}\,$. \vp{2mm}

\noi In the strong magnetic field $\,{\mathcal B}\,$ (or very
small mass $\,m$), Lagrangian (33) with  $\,{\mathcal E} = 0\,$
becomes \bea\hp{-9mm} L = \frac{e\hp{.3mm} {\mathcal B}}{2}
(\dot{x}_1\, x_2 - \dot{x}_2 \, x_1 ). \label{50} \eea The
corresponding canonical momentum is $\,p_1 =\frac{\partial
L}{\partial \dot{x}_1} =\frac{e\hp{.3mm} {\mathcal B}}{2}\, x_2\,$
and \bea\nn \hp{-7mm} [\hp{.3mm}\hat{x}_1 , \hat{p}_1\hp{.3mm}]  =
\Big[\hat{x}_1 , \frac{e\hp{.3mm} {\mathcal
B}}{2}\hp{.3mm}\hat{x}_2\Big] = i\hp{.3mm} \hbar\hp{.3mm} \Big( 1+
\frac{\theta\hp{.3mm} \sigma}{4} \Big).\eea  In this limit one
emerges NC configuration space with $\, [\hat{x}_1 , \hat{x}_2] =
i \hbar \frac{2}{e\hp{.3mm} {\mathcal B}} \big( 1+ \frac{\theta
\hp{.3mm} \sigma}{4} \big) .$ The spacing between energy levels
diverges like $\,- \frac{\hbar\hp{.5mm} e^2\hp{.2mm} {\mathcal
B}^2\hp{.2mm} \theta}{4 m}\,$ and the system practically lives in
the lowest level $\,- \frac{\hbar\hp{.5mm} e^2\hp{.2mm} {\mathcal
B}^2\hp{.2mm} \theta}{8\hp{.3mm} m} ,$ which differs from the
standard Landau lowest level $\,\frac{\hbar\hp{.4mm} e
\hp{.3mm}{\mathcal B}}{2\hp{.3mm} m}\,.$

\vp{2mm}

\noi At the end, it is worth noting that there are some other
commutation relations similar to (1), which by linear
transformations of canonical variables can be converted to the
usual Heisenberg algebra (2). This will be presented elsewhere.

\section*{Acknowledgments}

The work on this article was partially supported by the Ministry
of Science and Environmental Protection, Serbia, under contracts
No 1426 and No 1646.

%\bigskip

\end{document}